\documentclass[preprint,11pt,preprintnumbers,nofootinbib]{revtex4}
\usepackage{graphicx}
\usepackage{dcolumn}
\usepackage{bm}
\usepackage{amsfonts}

\begin{document}

\date{\today}
\title{On Non-Gaussianities in Multi-Field Inflation (N fields):\\
Bi- and Tri-spectra beyond Slow-Roll}
\author{Diana Battefeld$^{1,2)}$} 
\email{dbattefe(AT)princeton.edu}
\author{Thorsten Battefeld$^{2)}$}
\email{tbattefe(AT)princeton.edu}
\affiliation{1) Helsinki Institute of Physics, P.O. Box 64, FIN-00014 Helsinki, Finland}
\affiliation{2) Princeton University, Department of Physics, NJ 08544, USA}

\begin{abstract}
We compute analytic expressions for the non-linearity parameters characterizing the bi- and tri-spectrum of primordial curvature perturbations generated during an inflationary epoch of the early universe driven by an arbitrary number of fields. We assume neither slow roll nor a separable potential; instead, to compute Non-Gaussianities, we assume a separable Hubble parameter. We apply the formalism to an exact solvable toy-model and show under which conditions observably large non-Gaussianities are produced.    

\end{abstract}
\maketitle
\newpage

\tableofcontents

\section{Introduction}

The spectrum of temperature fluctuations in the cosmic microwave background radiation (CMBR) is commonly accepted to have been generated by the amplification of vacuum fluctuations in light fields during an inflationary epoch of the early universe. 
Observations of the CMBR \cite{COBE,Spergel:2006hy,Maxima,Boomerang,Komatsu:2008hk} are in agreement with the prediction  based on the simplest, single-field inflationary models of a nearly scale invariant, highly Gaussian spectrum of adiabatic cosmological perturbations, see i.e.~\cite{Lyth:1998xn,Bassett:2005xm} for reviews; 
the validity of the inflationary framework has withstood the tests of time since its inception in $1980$. Our interest in density perturbations, commonly characterized by the comoving curvature perturbation $\zeta$, stems from the fact that one can differentiate between models by considering higher order correlation functions, which will be probed in upcoming experiments \cite{Komatsu:2009kd}, such as PLANCK \cite{Planck}. 

The non linearity parameters $f_{NL}$ \cite{Komatsu:2001rj}, characterizing the three-point correlation function, and $\tau_{NL}$ \cite{Seery:2006js,Byrnes:2006vq}, characterizing the four-point correlation function, are widely used to estimate non-Gaussianities (NG). 
Even though primordial NG, i.e.~a non-zero three-point function, have not yet been observed, analysis of the 5-year WMAP data provides strong indications for their presence, $-9<f_{NL}^{local}<111$ at $95\%$CL \cite{Komatsu:2008hk} or $-4 < f_{NL}^{local} < 80$ at $95\%$ CL in the  re-analysis by K. Smith et.al.~\cite{Smith:2009jr}. 
The recently launched PLANCK satellite \cite{Planck} will be sensitive enough to detect NG in the next years with $\Delta f_{NL}\sim \pm 5$, if they are indeed present at the currently expected level; further, bounds on $\tau_{NL}$ will be considerably improved.

Simple single-field models have typically tiny non-Gaussianities  \cite{Maldacena:2002vr,Acquaviva:2002ud,Creminelli:2003iq,Babich:2004gb,Bartolo:2004if,Seery:2005wm,Seery:2008ax}. However,  more complicated models warrant the possibility of allowing for larger NG within the inflationary framework; examples are features in the potential \cite{Chen:2006xjb,Chen:2008wn}, preheating \cite{Enqvist:2004ey,Jokinen:2005by,Barnaby:2006cq,Barnaby:2006km}, modulated preheating \cite{Dvali:2003em,Battefeld:2007st,Suyama:2007bg,Byrnes:2008zz}, the presence of additional light fields, and thus isocurvature perturbations, as in the curvaton-scenario \cite{Linde:1996gt,Bartolo:2001cw,Bernardeau:2002jf,Linde:2005yw,Boubekeur:2005fj,Sasaki:2006kq,Malik:2006pm,Kawasaki:2008sn,Huang:2008zj,Hikage:2008sk,Langlois:2008vk,Li:2008fma,Ichikawa:2008iq,Dutta:2008if}, trapped fields \cite{Suyama:2008nt}, DBI-inflation/k-inflation \cite{Alishahiha:2004eh,Chen:2006nt,Huang:2006eh,Arroja:2008ga,Arroja:2008yy,Gao:2009gd,Cai:2009hw,Khoury:2008wj,Arroja:2009pd,Chen:2009bc,Mizuno:2009mv,Gao:2009at,Mizuno:2009cv,Langlois:2009ej,Langlois:2008qf,Langlois:2008wt,RenauxPetel:2009sj}, 
nonlocal inflation \cite{Barnaby:2007yb,Barnaby:2008fk}, among others. Alternative proposals, such as the new-ekpyrotic scenario \cite{Buchbinder:2007ad,Creminelli:2007aq,Lehners:2007ac}, have usually quite strong NG signals too \cite{Koyama:2007if,Buchbinder:2007at,Khoury:2008wj,Lehners:2009ja,Lehners:2007wc,Lehners:2008my}, see \cite{Lehners:2008vx} for a review.

 In this article, we further investigate the possibility to generate NG in multi-field inflationary models. The generation of NG in multi-field models was considered in \cite{Bernardeau:2002jy,Alabidi:2005qi,Lyth:2005fi,Lyth:2005du,Seery:2005gb,Battefeld:2006sz,Seery:2006vu,Byrnes:2007tm,Byrnes:2006vq,Yokoyama:2007uu,Yokoyama:2007dw} (see also \cite{Salem:2005nd,Lyth:2005qk,Alabidi:2006wa,Alabidi:2006hg,Sasaki:2008uc,Byrnes:2008zy,Byrnes:2008wi,Naruko:2008sq} for hybrid/multi-brid models, and \cite{Huang:2009xa,Gao:2008dt,Battefeld:2007en,Choi:2007su,Misra:2007cq,Misra:2008tx} for related work), where it was found that the non-linearity parameters are usually slow roll suppressed\footnote{In some cases large NG are possible during slow roll even in simple models, see i.e.~\cite{Byrnes:2008wi} for fine tuning initial conditions, \cite{Rodriguez:2008hy,Cogollo:2008bi} for contributions of loops or \cite{Huang:2009vk} for effects related to geometric quantities of the hyper-surface at which inflation ends.}. 
The presence of isocurvature modes in multi-field models \cite{Gordon:2000hv,Malik:2005cy,Malik:2006ir} can cause Fourier-modes of the curvature perturbation $\zeta_k$ to evolve even after horizon exit, provided that the trajectory in field space is curved. Hence, NG can be sourced \cite{Bernardeau:2002jy}, but a sharp turn, and thus violation of the slow-roll conditions, is usually required to give rise to large NG. A computation of $f_{NL}$ involving two fields, a separable potential, and the slow-roll approximation was made by Vernizzi and Wands \cite{Vernizzi:2006ve}, which was later extended to an arbitrary number of fields in \cite{Battefeld:2006sz} and the tri-spectrum ($\tau_{NL}$) in \cite{Seery:2006vu}. 
However,  slow roll can be temporarily violated during inflation, i.e.~if a bump in the potential is encountered, if fields start to decay during inflation as in staggered/cascade inflation \cite{Becker:2005sg,Ashoorioon:2006wc,Battefeld:2008py,Battefeld:2008ur,Battefeld:2008qg}, and it is necessarily violated at the end of inflation and during (p)re-heating. Unfortunately, almost all analytic studies based on the (non-linear) $\delta N$-formalism \cite{Lyth:2004gb,Lyth:2005fi}\footnote{This formalism is a crucial extension of the linear $\delta N$-formalism \cite{Starobinski,Sasaki:1995aw}, and needed to deal with higher order correlation functions.} are based on the slow roll approximation, rendering them unapplicable in these cases (see however \cite{Yokoyama:2007dw}); other computational techniques that do not involve the slow roll approximation are much more involved and analytic results for higher order correlation functions become unfeasible.

Recently, Byrnes and Tasinato \cite{Byrnes:2009qy} found a class of inflationary models that are amendable to an analytic computation of NG within the $\delta N$-formalism, without imposing the slow-roll or the horizon-crossing approximation, but requiring a separable Hubble parameter $H=\sum_k H_k(\varphi_k)$. They focused on two-field models with canonical kinetic terms and computed an expression for $f_{NL}$. 
In this paper, we generalize this framework to an arbitrary number of fields and provide expressions for $f_{NL}$ and $\tau_{NL}$. We demonstrate the applicability of the formalism in a concrete toy model with up to six inflaton fields that allows for the generation of large (but negative $f_{NL}<0$) NG towards the end of inflation. 

Due to the assumption of a separable Hubble parameter, the potential necessarily contains cross couplings between the inflatons. Thus, simple models of assisted inflation \cite{Liddle:1998jc,Malik:1998gy,Kanti:1999vt,Kanti:1999ie} do not fall into the class of models amendable to our treatment. However, the absence of cross couplings does in fact constitute fine tuning, and more realistic models of multi-field inflation in string theory are closer to the ones we consider. Furthermore, the presence of cross couplings should render the framework at hand applicable to an analytic computation of NG during (p)reheating, a project we plan to come back to in the future. 

The concrete outline of this article is as follows: After introducing the setup in Sec.~\ref{sec:setup}, we review briefly the $\delta N$-formalism in Sec.~\ref{sec:deltaN} and show how expressions for the power-spectrum, the bi-spectrum and the tri-spectrum may be obtained, Sec.~\ref{sec:power}-\ref{sec:taunl}. The needed derivatives of the volume expansion rate are computed in Sec.~\ref{sec:derivativesofn}, which are then used to provide analytic expressions for the power-spectrum, $f_{NL}$ and $\tau_{NL}$ in Sec.~\ref{sec:analyticpower}-\ref{sec:analytictaunl}. These are the main results of this paper. After a short discussion in Sec.~\ref{sec:discussion}, we provide a concrete example in Sec.~\ref{sec:examples}: a two-field model as introduced in \cite{Byrnes:2009qy} (Sec.~\ref{caseN=2}), and a case with up to six inflatons (Sec.~\ref{casengeneral}). We conclude in Sec.~\ref{sec:conclusion}.

\section{The Setup \label{sec:setup}}
The statistical information about primordial perturbations can be extracted from correlation functions of the curvature perturbation $\zeta$, which are imprinted onto the temperature fluctuations of the CMBR after inflation. 
Our goal is to derive expressions for the nonlinearity parameters $f_{NL}$ and $\tau_{NL}$ that characterize the magnitude of the three- and four-point correlation functions (bi- and tri-spectrum) of  $\zeta$ by use of the (non-linear) $\delta N$-formalism  \cite{Lyth:2004gb,Lyth:2005fi}  in multi-field inflationary models, with an arbitrary number of fields, without imposing slow roll, but assuming a separable Hubble parameter \cite{Salopek:1990jq}
\begin{eqnarray}
H=\sum_{k=1}^{\mathcal{N}}H_k(\varphi_{k})\,. \label{seperableH}
\end{eqnarray}
 As such, our treatment extends the work of Byrnes and Tasinato \cite{Byrnes:2009qy} who computed $f_{NL}$ under the same assumption in two-field models. Technically, \cite{Byrnes:2009qy} parallels the work of Vernizzi and Wands \cite{Vernizzi:2006ve}, who computed a general expression for $f_{NL}$ for two fields by assuming slow roll and a separable potential $V=\sum_k V_k(\varphi_k)$. The latter work was extended in \cite{Battefeld:2006sz} to an arbitrary number of fields and in \cite{Seery:2005gb} to the tri-spectrum. Here, we follow  \cite{Battefeld:2006sz,Seery:2005gb} closely. 

We focus on inflation driven by $\mathcal N$ scalar fields with the action
\begin{eqnarray}
S=\frac{1}{2}\int d^4x\sqrt{-g}\left(\frac{1}{2}\sum_{k=1}^{\mathcal{N}}\partial^{\mu}\varphi_k\partial_{\mu}\varphi_k+V(\varphi_1,\varphi_2,...)\right)\,.
\end{eqnarray}
Here and in the following we set the reduced Planck mass equal to one, $m_p=(8\pi G)^{-1/2}\equiv 1$; all sums run from $1$ to $\mathcal{N}$ unless specified otherwise.
Focussing on a homogeneous universe, the Friedmann and Klein-Gordon equations can be written as a first order Hamilton-Jacobi system \cite{Salopek:1990jq}
\begin{eqnarray}
H^2&=&\frac{1}{3}V+\frac{2}{3}\sum_k \left(\frac{\partial H}{\partial \varphi_k}\right)^2\,,\label{HJ1}\\
\dot{\varphi}_k&=&-2\frac{\partial H}{\partial \varphi_k}\,,\label{HJ2}
\end{eqnarray}
where a dot denotes a derivative with respect to cosmic time and $H=\dot{a}/a$. These equations are exact, namely, they do not rely on the slow roll approximation. Because (\ref{HJ2}) decouples if the Hubble parameter is separable, we specifically focus on this case. 
Inflation takes place as long as the Hubble slow-evolution parameter remains small,
\begin{eqnarray}
\epsilon\equiv -\frac{\dot{H}}{H^2} \ll 1\,.
\end{eqnarray}
In analogy to the slow roll parameters, it is useful to define
\begin{eqnarray}
\delta_k&\equiv &\left(\frac{H_{,k}}{H}\right)^2\,,\label{correspondingslowroll1}\label{deltak}\\
\gamma_k&\equiv &\frac{H_{,kk}}{H}\,,\label{correspondingslowroll2}\label{gammak}\\
\beta_k&\equiv &\frac{H_{,kkk}}{H}\frac{H_{,k}}{H}\,,\label{correspondingslowroll3}\label{betak}
\end{eqnarray}
where we used the short hand notation $H_{,k}\equiv \partial H/\partial\varphi_k$ \footnote{In the following computations, we assume $H_{,k}>0$; results, such as (\ref{fnle})-(\ref{tau4e}) are also valid for $H_{,k}<0$. Note that the ambiguity in the sign of $\sqrt{\delta_k}$ lead to a sign mistake for $f_{NL}$ in early versions of \cite{Byrnes:2009qy}. In this paper, all expressions have been correct to our knowledge, but Fig.~\ref{pic:NG_Byrnes}-\ref{pic:NG_Multi} contain wrong labels in the JCAP version and v1 and v2 on the arxiv (it should have been $f_{NL}^{(4)}$ instead of $-f_{NL}^{(4)}6/5$). The labels are corrected in this version v3. We thank C.Byrnes and G.Tasinato for alerting us to this mistake.}. Note the absence of any mixed derivatives due to the Ansatz (\ref{seperableH}). The universe inflates as long as
\begin{eqnarray}
\delta\equiv \sum_{k}\delta_k \ll \frac{1}{2}\,,\label{correspondingslowroll4}
\end{eqnarray}
because $\epsilon =2\delta$; however, neither $\gamma_k$ nor $\beta_k$ are required to be small for inflation to take place.

\section{The $\delta N$-formalism \label{sec:deltaN}}
The $\delta N$-formalism goes back to Starobinski \cite{Starobinski}, was extended by Sasaki and Stewart in \cite{Sasaki:1995aw} and generalized to higher orders in  \cite{Lyth:2004gb,Lyth:2005fi} (see also \cite{Vernizzi:2006ve,Seery:2005gb,Allen:2005ye} for related work \footnote{The separate universe formalism put forward by Rigopoulos and Shellard in e.g. \cite{Rigopoulos:2003ak} is equivalent to the $\delta N$-formalism.}). To employ the $\delta N$-formalism and compute correlation functions of $\zeta$, we need to evaluate the unperturbed volume expansion rate from an initially flat hypersurface at $t_*$ to a final uniform density hypersurface at $t_c$
\begin{eqnarray}
N(t_c,t_*)\equiv\int^{t_c}_{t_*} H\,dt\,.
\end{eqnarray}
In the following, all integrals are assumed to run over values from $t_*$ to $t_c$ if not specified otherwise. Since different trajectories in field space can provide the same homogeneous expansion rate, we define, in analogy to the $\mathcal{N}-1$ integrals of motion in slow roll inflation \cite{Battefeld:2006sz},
\begin{eqnarray}
C_k\equiv -\int\frac{d\varphi_k}{H_{,k}}+\int{\frac{d\varphi_{k+1}}{H_{,{k+1}}}}\,, \label{C}
\end{eqnarray}
for $k=1\dots \mathcal{N}-1$. These quantities are conserved during inflation and can be used to discriminate between different trajectories \footnote{The definition in (\ref{C}) is not unique and sets of (independent) linear combinations of the $C_k$'s may be used alternatively \cite{Seery:2006js}.}.
The perturbation of the expansion rate, $\delta N$, is identical  to the curvature perturbation $\zeta$, 
\begin{eqnarray}
 \zeta&=&\delta N =\sum_kN_{,k}\delta\varphi_{k}^*+\frac{1}{2}\sum_{kl}N_{,kl}\delta\varphi_{,k}^*\delta\varphi_{l}^*+...\,,\label{zeta}
 \end{eqnarray} 
which is valid to any order in perturbation theory \cite{Lyth:2004gb,Lyth:2005fi}. If a quantity is to be evaluated at $t_*$ or $t_c$ we denote it by a superscript, i.e. $H_{,k}^*=\partial H/\partial\varphi_k|_{t_*}$ \footnote{The derivatives of the volume expansion rate with respect to $\varphi_k^*$ depend on both, $t_c$ and $t_*$, so no superscript is used in (\ref{def:derivativeofN}). To avoid confusion with summation indices, we use  $k,l,m,\dots$ exclusively for indices that run from $1$ to $\mathcal{N}$ and $i,j$ for indices that run from $1$ to $4$.}. For simplicity we use the short hand notation 
\begin{eqnarray}
N_{,k}&\equiv&\frac{\partial N}{\partial\varphi_{k}^*}\,, \label{def:derivativeofN}\\
N_{,kl}&\equiv&\frac{\partial^2 N}{\partial\varphi_{k}^*\partial\varphi_{l}^*}\,,\\
\nonumber &\vdots&
\end{eqnarray}
With the Ansatz (\ref{seperableH}), we can simplify the volume expansion rate to \cite{Byrnes:2009qy}
\begin{eqnarray}
N(t_c,t_*)=-\frac{1}{2}\sum_{k=1}^{\mathcal N}\int_{\varphi_k^*}^{\varphi_k^c}\frac{H_k}{H_{,k}}d\varphi_k\,. \label{N}
\end{eqnarray}
Note that we did not assume slow roll, but $H$ needs to be separable in order for (\ref{N}) to hold \footnote{If the slow roll approximation is made and a separable potential is used, one can derive \cite{Lyth:1998xn} $N(t_c,t_*)=-\int_*^c \sum_{k=1}^{\mathcal N}(V_k/ V_k^\prime) d\varphi_k$ which also enables the analytic computation of non-linearity parameters within the $\delta N$-formalism.}.

\subsection{The power-spectrum \label{sec:power}}
The power-spectrum of $\zeta$ is defined in terms of the two-point correlation function as
\begin{eqnarray}
{<\zeta_{\bf{k_1}}\zeta_{\bf{k_2}}>}&\equiv&{(2\pi)^3\delta^3(\bf{k_1}+\bf{k_2})}\frac{2\pi^2}{k_1^3}\mathcal P_{\zeta}(k_{1})\,,
\end{eqnarray}
where $\zeta_{\bf{k}}$ denotes a Fourier mode of the curvature perturbation.
Using (\ref{N}), the power-spectrum can be related to the derivatives of the volume expansion rate,
\begin{eqnarray}
\mathcal{P}_{\zeta}=\sum_{k} N^2_{,k}\mathcal P^*\,,\label{powersp}
\end{eqnarray}
and the scalar spectral index and the tensor to scalar ratio become \cite{Vernizzi:2006ve}
\begin{eqnarray}
n_{s}-1&\equiv& \frac{\partial \ln \mathcal{P}_\zeta}{\partial \ln k}=-2\epsilon^*+\frac{2}{H^*}\frac{\sum_{kl=1}^{\mathcal N}\dot{\varphi}_{k}^*N_{,kl}N_{,l}}{\sum_{m=1}^{\mathcal N}N^2_{,m}}\,,\label{ns}\\
r&\equiv& \frac{\mathcal{P}_T}{\mathcal{P}_\zeta}=\frac{8\mathcal{P}^*}{\mathcal{P}_\zeta}\,,
\end{eqnarray}
where $\epsilon^*\equiv-(\dot{H}/H^2)^*$ and $\mathcal{P}^*=k^3P^*(k)/(2\pi^2)=(H^*)^{2}/(4\pi^2)$; these  have to conform to the 
COBE normalization $\mathcal{P}_\zeta=(2.41\pm 0.11) \times 10^{-9}$ \cite{Komatsu:2008hk} as well as  $n_s= 0.960 \pm 0.013$ from the WMAP $5$-year data analysis \cite{Komatsu:2008hk} and $r<0.22\; (95\%\;\mbox{CL})$ combining WMAP5, BAO and SN data \cite{Komatsu:2008hk}. In order to compute these observable quantities, we need to evaluate the derivatives of $N$ with respect to $\varphi_k^*$, see Sec.~\ref{sec:derivativesofn}.

\subsection{The bi-spectrum and $f_{NL}$ \label{sec:fnl}}
The bi-spectrum of $\zeta$ is defined in terms of the three-point correlation function as 
\begin{eqnarray}
{<\zeta_{\bf{k_1}}\zeta_{\bf{k_2}}\zeta_{\bf{k_3}}>}&\equiv&{(2\pi)^3\delta^3(\bf{k_1}+\bf{k_2}+\bf{k_3})}\mathcal B_{\zeta}(k_{1},k_{2},k_{3})\,.
\end{eqnarray}
A measure of its magnitude is the non-linearity parameter $f_{NL}$, defined as\footnote{In v3 of this article we switched to the sign convention of Komatsu and Spergel \cite{Komatsu:2001rj,Komatsu:2008hk} for $f_{NL}$. v1 and v2 as well as the JCAP version used the opposite one by Vernizzi and Wands \cite{Vernizzi:2006ve}.}
\begin{eqnarray}
\frac{6}{5}f_{NL}&\equiv& \frac{k_1^3k_2^3k_3^3}{k_1^3+k_2^3+k_3^3}\frac{\mathcal{B}_\zeta}{4\pi^2 \mathcal{P}_\zeta^2} \,.
\end{eqnarray}
Using (\ref{N}), one can relate the non-linearity parameter $f_{NL}$ to the derivatives of the expansion rate $N$ with respect to the field values as \cite{Vernizzi:2006ve,Seery:2005gb},
\begin{eqnarray}
\frac{6}{5}f_{NL}&=&\frac{r}{16}(1+f)+\frac{\sum_{kl}N_{,k}N_{,l}N_{,kl}}{(\sum_{k}N^2_{,k})^2}\label{fnl}\\ 
&\equiv & \frac{r}{16}(1+f)-\frac{6}{5}f_{NL}^{(4)}\,.\label{r}
\end{eqnarray}
Here, $f$ includes the contingency of $f_{NL}$ on the shape of the momentum triangle, $0\leq f\leq 5/6$ \cite{Maldacena:2002vr,Vernizzi:2006ve}; 
$f$ is maximal for an equilateral triangle and  minimal for a triangle where two sides are much longer than the third \cite{Maldacena:2002vr}. Since the tensor:scalar ratio $r$ is smaller than one,  it is evident that the first term in (\ref{fnl}) is suppressed. The WMAP$5$ data \cite{Komatsu:2008hk} currently yields the bound $-9<f_{NL}^{local}<111$ at $95\%$CL or $-4 < f_{NL}^{local} < 80$ at $95\%$ CL in the  re-analysis by K. Smith et.al.~\cite{Smith:2009jr}, which will be improved considerably in the near future by PLANCK \cite{Planck}. Nevertheless, unless $f_{NL}>1$, non-Gaussianities are unlikely to be ever detected. For this reason, from here on, we focus on the second term in (\ref{fnl}), which provides a momentum independent contribution to $f_{NL}$. 

\subsection{The tri-spectrum and $\tau_{NL}$\label{sec:taunl}}
The tri-spectrum of $\zeta$ is defined in terms of the four-point correlation function as 
\begin{eqnarray}
{<\zeta_{\bf{k_1}}\zeta_{\bf{k_2}}\zeta_{\bf{k_3}}\zeta_{\bf{k_4}}>}&\equiv&{(2\pi)^3\delta^3(\bf{k_1}+\bf{k_2}+\bf{k_3}+\bf{k_4})}T_{\zeta}({\bf{k}}_1,{\bf{k}}_2,{\bf{k}}_3,{\bf{k}}_4)\,,
\end{eqnarray}
where only the connected part of the correlator is considered \cite{Okamoto:2002ik,Kogo:2006kh}.
To estimate the magnitude of $T_{\zeta}$, it is common  to define $\tau_{NL}$ by 
\begin{eqnarray}
T_\zeta\equiv\frac{1}{2}\tau_{NL}\bigg(\sum_{\alpha} N^2_{,\alpha}\bigg)^3[P^*(k_1)P^*(k_2)P^*(k_{14})+23 \mbox{ permutations}]\,,\label{T}
\end{eqnarray}
where ${\bf{k}}_{ij}={\bf{k}}_{i}+{\bf{k}}_{j}$.
Defining
\begin{eqnarray}
 T\equiv\sum_{i<j}\sum_{s\neq ij}k_i^3k_j^3\left(k_{is}^{-3}+k_{js}^{-3}\right)\,,
\end{eqnarray}
the momentum dependence in (\ref{T}) can be made explicit 
\cite{Seery:2006vu} 
\begin{eqnarray}
T_{\zeta}=\frac{4\pi^6}{\Pi_ik^3_i}\tau_{NL}\bigg(\sum_{k}(N^*_{,k })^2\bigg)^3(\mathcal{P}^*)^3 T\,.
\end{eqnarray}
$\tau_{NL}$ contains four tree-level components $\Delta \tau_{NL}^{(i)}$, $i=1\dots 4$. One component can be shown to be bounded from above by the tensor:scalar ratio \cite{Seery:2006vu}, $|\Delta\tau_{NL}^{(1)}|\lesssim r/50$, which is too small to be observable and thus of no interest to us. The remaining three contributions are \footnote{In the notation of Byrnes \cite{Byrnes:2006vq,Byrnes:2009qy} $\Delta\tau_{NL}^{(3)}=\tau_{NL}^{(\mbox{\tiny Byrnes})}$ 
and $\Delta\tau_{NL}^{(4)}=g_{NL}^{(\mbox{\tiny Byrnes})} 27{T}/(25\sum_ik_i^3)$.} \cite{Seery:2005gb} (see also \cite{Byrnes:2006vq,Alabidi:2005qi})
\begin{eqnarray}
\Delta \tau_{NL}^{(2)}&=&\frac{\sum_{kl}\dot{\varphi}_{k}^*N_{,kl}N_{,l}}{\left(\sum_lN^2_{,l}\right)^2}\frac{\mathcal K}{4H^*}\,,\label{tau2}\\
\Delta \tau_{NL}^{(3)}&=&\frac{\sum_{klm}N_{,kl}N_{,ml}N_{,k}N_{,m}}{\left(\sum_l N^2_{,l}\right)^3}\,,\label{tau3}\\
\Delta \tau_{NL}^{(4)}&=&2\frac{\sum_{klm}N_{,klm}N_{,k}N_{,l}N_{,m}}{\left(\sum_lN^2_{,l}\right)^3}\frac{\sum_ik^3_i}{T}\,,\label{tau4}
\end{eqnarray}
where $\mathcal K$ is defined as
\begin{eqnarray}
\mathcal K=\frac{1}{T}\sum_{perms}\frac{k_1^3}{k^3_{12}}\mathcal M(k_{12},k_{3},k_{4})\,,
\end{eqnarray}
and
\begin{eqnarray}
\mathcal M(k_1,k_2,k_3)\equiv-k_1k_2^2-4\frac{k^2_2k^2_3}{k_1+k_2+k_3+k_4}+\frac{1}{2}k_1^3+\frac{k_2^2k_3^2(k_2-k_3)}{(k_1+k_2+k_3+k_4)^2}\,.
\end{eqnarray}

\section{The Derivatives of $N$ \label{sec:derivativesofn}}
To evaluate the non-linearty parameters $f_{NL}$ and $\tau_{NL}$ we need to compute the derivatives of the volume expansion rate with respect to the fields. In this section, we follow \cite{Battefeld:2006sz} and \cite{Seery:2006vu} closely. Based on (\ref{N}), the total differential of $N$ reads
\begin{eqnarray}
dN=\sum_{k=1}^{\mathcal N}\frac{1}{2}\left[\left(\frac{H_{k}}{H_{,{k}}}\right)^*-\sum_{l=1}^{\mathcal{N}}\frac{\partial\varphi_{l}^c}{\partial\varphi_{k}^*}\left(\frac{H_{l}}{H_{,{l}}}\right)^c\right]d\varphi_{k}^*\,.\label{dN}
\end{eqnarray}
Using the integrals of motion $C_k$ in (\ref{C}), we can relate $d\varphi_{k}^c$ and $d\varphi_{k}^*$
\begin{eqnarray}
d\varphi_{m}^c=\sum^{\mathcal N-1}_{l=1}\frac{\partial\varphi_{m}^c}{\partial C_l}\left(\sum_{k=1}^{\mathcal N}\frac{\partial C_l}{\partial\varphi_{k}^*}d\varphi_{k}^*\right)\,,\label{dphij}
\end{eqnarray}
where
\begin{eqnarray}
\frac{\partial C_l}{\partial\varphi_{k}^*}=\frac{1}{H_{,{k}}^*}\left(\delta_{lk-1}-\delta_{lk}\right)\,.\label{delta}
\end{eqnarray}
We want to evaluate NG at $t_c$ where the condition $\rho=\mbox{const}$ holds; in order to evaluate this condition, we first eliminate $\partial\varphi_{j}^c/\partial C_i$ in favor of $\partial\varphi_{1}^c/\partial C_i$. Hence we define
\begin{eqnarray}
\tilde{C_k}&\equiv&\sum^{k-1}_{l=1}C_l\\
&=&-\int\frac{d\varphi_1}{H_{,1}}+\int\frac{d\varphi_k}{H_{,k}}\,,
\end{eqnarray}
and differentiate with respect to $C_k$,
\begin{eqnarray}
\frac{\partial\tilde C_l}{\partial C_k}=-\frac{\partial\varphi_{1}^c}{\partial C_k}\frac{1}{H_{\prime{1}}^c}+\frac{\partial\varphi_{l}^c}{\partial C_k}\frac{1}{H_{,{l}}^c}\,,
\end{eqnarray}
so that
\begin{eqnarray}
\frac{\partial\varphi_{l}^c}{\partial C_k}=\frac{H_{,l}^c}{H_{,{1}}^c}\frac{\partial\varphi_{1}^c}{\partial C_k}+H_{,{1}}^c\Theta_{kl}\,,
\end{eqnarray}
where we defined
\begin{eqnarray}
\mathbf{\Theta}_{kl}=
\left\{\begin{array}{ll}
1,&\textrm{if k} \leq l-1\\
0,&\textrm{if k} >l-1\,.
\end{array}\right.
\end{eqnarray}
Inserting this into the derivative (with respect to $C_k$) of the $\rho=\mbox{const}$ condition, that is into
\begin{eqnarray}
0=\sum_{l=1}^{\mathcal N}H_{,{l}}^c\frac{\partial \varphi_{l}^c}{\partial C_k}\,,
\end{eqnarray}
we obtain
\begin{eqnarray}
\frac{\partial\varphi_{l}^c}{\partial C_k}=-\left[H_{,l}\frac{\sum^{\mathcal N}_{m=k+1}H_{,m}^2}{\sum_{m=1}^{\mathcal N}H_{,{m}}^2}\right]^c+H_{,l}^c\Theta_{kl}\,.\label{thetaeqn}
\end{eqnarray}
Using (\ref{thetaeqn}), (\ref{dphij}) and (\ref{delta}) we arrive at
\begin{eqnarray}
\frac{\partial\varphi_{l}^c}{\partial\varphi^*_{k}}&=&-\frac{H_{,{l}}^c}{H_{,{k}}^*}\left[\frac{H_{,{k}}^2}{\sum_{m=1}^{\mathcal N}H_{,m}^2}-\delta_{lk}\right]^c\\
&=&-\frac{H^c}{H^*}\sqrt{\frac{\delta_{l}^c}{\delta_{k}^*}}\left[\frac{\delta_{k}^c}{\delta^c}-\delta_{lk}\right]\,. \label{diffphicphi*}
\end{eqnarray}
To write the derivatives of $N$, it is convenient to define
\begin{eqnarray}
E_{km}&\equiv&\frac{\delta_k^c}{\delta^c}-\delta_{mk}\,,\label{E}\\
Y_k&\equiv&\delta_k^c\left(1-\frac{2\gamma_k^c}{\delta^c}\right)\,,\label{Y}\\
X_k&\equiv&2\left[\gamma_k^cY_k-\frac{\delta_k^c\beta_k^c}{\delta^c}\right]\,,\label{X}
\end{eqnarray}
with $\delta_k$, $\gamma_k$ and $\beta_k$ from (\ref{deltak})-(\ref{betak}). Differentiating $\delta_k$ and $\beta_k$ yields
\begin{eqnarray}
\frac{\partial\delta_{l}^c}{\partial\varphi_{m}^c}&=&2\left(\gamma_{l}\delta_{lm}-\sqrt{\delta_{l}\delta_{m}}\right)^c\sqrt{\delta_{l}^c}\,,\label{der1}\\
\frac{\partial\delta^c}{\partial\varphi_{m}^c}&=&2\sqrt{\delta_{m}^c}\left(\gamma_{m}-\delta\right)^c\,,\label{der2}\\
\frac{\partial\gamma_{m}^c}{\partial\varphi_{l}^c}&=&\sqrt{\delta_{l}^c}\left(\delta_{ml}\frac{\beta_{m}}{\delta_{m}}-\gamma_{m}\right)^c\,.\label{der3}
\end{eqnarray}
Further, the derivatives of $E_{km}$ and $Y_{k}$ read
\begin{eqnarray}
\frac{\partial E_{km}}{\partial \varphi_{l}^c}&=&-2\left(\frac{\sqrt{\delta_{l}}\gamma_{l}}{\delta}\right)^c E_{kl}\,,\\
\frac{\partial Y_{k}}{\partial \varphi_{l}^c}&=&\frac{\delta_{kl}}{\sqrt{\delta_{l}^c}}X_{l}-\frac{2}{\sqrt{\delta_{l}^c}}\left(\gamma_{k}\frac{\delta_{k}}{\delta}Y_{l}+\delta_{l}Y_{k}\right)^c\,,
\end{eqnarray}
where we have used equations (\ref{E})-(\ref{X}) and (\ref{der1})-(\ref{der3}).

By means of equation (\ref{dN}), along with (\ref{diffphicphi*}) and the above, we obtain the first derivatives of the expansion rate with respect to $\varphi_k^*$ as
\begin{eqnarray}
N_{,k}=\frac{1}{2}\frac{u_k}{\sqrt{\delta_{k}^*}}\,, \label{firstN}
\end{eqnarray}
where we introduced
\begin{eqnarray}
u_k&\equiv&\frac{H_{k}^*+Z_{k}^c}{H^*}\,,\\
Z_{k}^c&\equiv& H^c\frac{\delta_k^c}{\delta^c}-H_k^c\,.
\end{eqnarray}
The second derivatives of $N$ become
\begin{eqnarray}
N_{,kl}=\frac{1}{2}\left[\delta_{kl}\left(1-\frac{\gamma_{k}^*}{\delta_{k}^*}u_k\right)+\frac{1}{\sqrt{\delta_{k}^*}}\frac{1}{H^*}\frac{\partial Z_{k}^c}{\partial \varphi_{l}^*}\right]\,,\label{secondN}
\end{eqnarray}
where
\begin{eqnarray}
\frac{\partial Z_{k}^c}{\partial\varphi_{l}^*}&=&\frac{H^*}{\sqrt{\delta_{l}^*}}A_{kl}\,,\label{derZkc}
\end{eqnarray}
and we have defined the symmetric matrix,
\begin{eqnarray}
A_{kl}&\equiv &-\left(\frac{H^c}{H^{*}}\right)^2\bigg(\sum_{m}E_{km}E_{lm}Y_{m}\bigg)^c\,.
\end{eqnarray}
The third derivatives become,
\begin{eqnarray}
N_{,klm}&=&\frac{\delta_{klm}}{2\sqrt{\delta_{k}^*}}\left[\frac{2(\gamma^*_k)^{2}-\beta_{k}^*}{\delta_{k}^*}u_k-\gamma_{k}^*\right]-\frac{\gamma_{k}^*}{2\delta_{k}^*H^*}\left(\delta_{km}\frac{\partial Z_{k}^c}{\partial\varphi_{l}^*}+\delta_{kl}\frac{\partial Z_{k}^c}{\partial\varphi_{m}^*}\right)\\
\nonumber &&+\frac{1}{2}\frac{1}{\sqrt{\delta_{k}^*}H^*}\frac{\partial^2 Z_{k}^c}{\partial\varphi_{l}^*\partial\varphi_{m}^*}\,,\label{thirdN}
\end{eqnarray}
where
\begin{eqnarray}
\frac{\partial^2Z_{k}^c}{\partial\varphi_{l}^*\partial\varphi_{m}^*}&=&-\delta_{lm}\frac{\gamma_{m}^*}{\delta_{m}^*}H^*A_{kl}+\frac{H^*}{\sqrt{\delta_{m}^*\delta_{l}^*}}A_{klm}\,,
\end{eqnarray}
and we defined the totally symmetric tensor 
\begin{eqnarray}
A_{klm}&\equiv&\left(\frac{H^{c}}{H^*}\right)^3\sum_{\alpha=1}^{\mathcal{N}}\Bigg[X_{\alpha}E_{k\alpha}E_{l\alpha}E_{m\alpha}+\sum_{\beta=1}^{\mathcal{N}}Y_{\beta}(Y_{\alpha}-\delta_{\alpha}^c)\\
\nonumber &&\times(E_{k\beta}E_{l\alpha}E_{m\alpha}+E_{k\alpha}E_{l\beta}E_{m\alpha}+E_{k\alpha}E_{l\alpha}E_{m\beta})\Bigg]\,.
\end{eqnarray}

\section{Analytic Expressions \label{sec:analyticexpressions}}
Equipped with $N_{,k}$ from (\ref{firstN}), $N_{,kl}$ from (\ref{secondN}) and $N_{,klm}$ from (\ref{thirdN}), we are now ready to compute observables such as the scalar spectral index as well as the momentum independent components of the non-linearity parameters $f_{NL}$ and $\tau_{NL}$.

\subsection{The power-spectrum \label{sec:analyticpower}}
Using the derivatives of the expansion rate  $N_{,k}$ and $N_{,kl}$ we can simplify the power-spectrum in (\ref{powersp}) to
\begin{eqnarray}
\mathcal P_{\zeta}=\frac{1}{4}\sum_k\frac{u_k^2}{\delta_{k}^*}\mathcal{P}^*\,,
\end{eqnarray}
and the scalar spectral index in (\ref{ns}) to
\begin{eqnarray}
n_{s}-1=4\delta^{*}-4\frac{\sum_k\left(1-\frac{\gamma_{k}^*u_k}{\delta_{k}^*}\right)u_k+\sum_{kl}\frac{u_l}{\delta_{l}^*}A_{kl}}{\sum_ n\frac{u_n^2}{\delta_{n}^*}}\,.\label{ns2}
\end{eqnarray}
This expression reduces to the ones in \cite{Byrnes:2009qy} for $\mathcal{N}=2$ and mimics a similar expression in the slow roll case with a  separable potential \cite{Battefeld:2006sz}; note however that the meaning of quantities such as $u_k$ as well as numerical factors differ from the slow roll case \footnote{In general, our assumption of a separable Hubble parameter excludes cases with a separable potential; hence neither case can be derived as the limit of the other.}.

\subsection{The bi-spectrum: $f_{NL}^{(4)}$\label{sec:analyticfnl}}
The general expression for $f_{NL}^{(4)}$   in (\ref{fnl}) can be simplified to
\begin{eqnarray}
-\frac{6}{5}f^{(4)}_{NL}=2\frac{\sum_{k}\frac{u_k^2}{\delta_{k}^*}\left(1-\frac{u_k\gamma_{k}^*}{\delta_{k}^*}\right)+\sum_{kl}\frac{u_ku_l}{\delta_{k}^*\delta_{l}^*}A_{kl}}{\left(\sum_{m}\frac{u_m^2}{\delta_{m}^*}\right)^2}\,. \label{fnle}
\end{eqnarray}
Again, this result reduces to the one found in \cite{Byrnes:2009qy} for two fields and mimics formally the slow roll result of \cite{Battefeld:2006sz}, except for different numerical factors.

\subsection{The tri-spectrum: $\Delta \tau_{NL}^{(i)}$, $i=2,3,4$ \label{sec:analytictaunl}}
The three parameters in (\ref{tau2})-(\ref{tau4}) can be written as
\begin{eqnarray}
\Delta\tau^{(2)}_{NL}&=&\frac{\mathcal K}{4}\frac{-2^3}{\left(\sum_n\frac{u_n^2}{\delta_{n}^*}\right)^2}\Bigg[\sum_k\left(1-\frac{\gamma_{k}^*}{\delta_{k}^*}{u_k}\right)u_k+\sum_{kl}A_{kl}\frac{u_l}{\delta_{l}^*}\Bigg]\,,\label{tau2e}\\
\Delta\tau^{(3)}_{NL}&=&\frac{2^2}{\left(\sum_n\frac{u_n^2}{\delta_{n}^*}\right)^3}\Bigg[\sum_m\left(1-\frac{\gamma_{m}^*}{\delta_{m}^*}{u_m}\right)^2\frac{u_m^2}{\delta_{m}^*}+2\sum_{lm}\left(1-\frac{\gamma_{m}^*}{\delta_{m}^*}{u_m}\right)A_{lm}\frac{u_lu_m}{\delta_{l}^*{\delta_{m}^*}}\label{tau3e}\\
\nonumber &&+\sum_{klm}\frac{A_{kl}A_{km}}{\delta_{k}^*}\frac{u_lu_m}{\delta_{l}^*\delta_{m}^*}\Bigg]\,,\\
\Delta\tau^{(4)}_{NL}&=&\frac{\sum_ik_i^3}{T}\frac{2^3}{\left(\sum_n\frac{u_n^2}{\delta_{n}^*}\right)^3}\Bigg[\sum_m\frac{u_m^3}{(\delta_{m}^*)^2}\left(\frac{2(\gamma_{m}^*)^2-\beta_m^*}{\delta_m^*}u_m-\gamma_{m}^*\right)-3\sum_{lm}\frac{u_lu_m^2}{\delta_{l}^*(\delta_{m}^*)^2}\gamma_{m}^*A_{ml}\label{tau4e}\\
\nonumber &&+\sum_{klm}\frac{u_ku_lu_m}{\delta_{k}^*\delta_{l}^*\delta_{m}^*}A_{klm}\Bigg]\,.
\end{eqnarray}
As expected, these expressions mimic formally the corresponding ones in \cite{Seery:2006vu}; however, we would like to emphasize, once again, that numerical factors as well as the definition and meaning of quantities differ.

\subsection{Discussion \label{sec:discussion}}
The expressions for $n_s$ in (\ref{ns2}), $f_{NL}^{(4)}$ in (\ref{fnle}) and $\tau_{NL}^{(i)}$, $i=2,3,4$ in (\ref{tau2e})-(\ref{tau4e}) are our main results. They are applicable for any set of scalar fields with canonical kinetic terms and potentials such that a separable Hubble parameter results. Since fields can evolve fast and couplings between the fields are included, an application to the end phase of inflation, including preheating, appears feasible.

Because $A_{kl}$ and $A_{klm}$ are symmetric in their indices, one can readily check that these terms only contribute to the non-linearity parameters if the fields are not multiples of each other, that is whenever the trajectory in field space is curved. This is expected, since in these instances isocurvature perturbations can influence the adiabatic mode.

The expressions for $f_{NL}$ and $\tau_{NL}^{(i)}$, $i=2,3,4$ mimic formally the corresponding ones in slow roll models in \cite{Battefeld:2006sz,Seery:2006vu}, but numerical factors as well as the definition of terms differ. In the case of two fields, we recover the results of \cite{Byrnes:2009qy} for $f_{NL}^{(4)}$.

\section{Example \label{sec:examples}}
We would like to show the applicability of the formalism using a simple example which allows for the generation of large (but negative, $f_{NL}<0$) NG. 
Consider $H=\sum_k H_k$ with
\begin{eqnarray}
H_k(\varphi_k)=\frac{H_0}{\mathcal{N}}\left(1-A_k e^{-\alpha_k\varphi_k}\right)\,,
\end{eqnarray}
with $A_k>0$. For $\mathcal{N}=2$, this example reduces to the second one of \cite{Byrnes:2009qy} (a sign mistake in \cite{Byrnes:2009qy} and wrong labels in Fig.~\ref{pic:NG_Byrnes}-\ref{pic:NG_Multi} of this paper are corrected in this version, v3.), so we can easily compare results. We first note that any $A_k\neq 1$ can be reabsorbed into a rescaling of $\varphi_k$.  Second, if $\alpha_k<0$, we can simply redefine $\varphi_k\rightarrow -\varphi_k$ and $\alpha_k\rightarrow -\alpha_k$. Hence, we can set $A_k\equiv 1$ and $\alpha_k>0$ for all $k$ without loss of generality, so that
\begin{eqnarray}
H=H_0\left(1-\sum_k\frac{1}{\mathcal{N}}e^{-\alpha_k\varphi_k}\right)\,. \label{Hexample1}
\end{eqnarray}
The Hamilton-Jacobi equations of motion in (\ref{HJ1}) and (\ref{HJ2}) are solved by
\begin{eqnarray}
a(\tau)&=&a_0e^{\tau}\prod_k\left(e^{\alpha_k\varphi_k^*}-\frac{2}{\mathcal{N}}\alpha_k^2\tau\right)^{1/(2\alpha_k^2)}\,,\\
\varphi_{k}(\tau)&=&\frac{1}{\alpha_k}\ln\left(e^{\alpha_k\varphi_k^*}-\frac{2}{\mathcal{N}}\alpha_k^2\tau\right)\,,
\end{eqnarray}
where we defined the dimensionless time $\tau\equiv H_0t$. We set $t_*=0$, so that $\varphi_k(0)=\varphi_k^*$.
The corresponding potential of the scalar fields is
\begin{eqnarray}
V=3H^2-2\sum_k H_{,k}^2\,.
\end{eqnarray} 
The potential has a wide plateau with $V\approx 3H_0^2$ that drops off steeply for $\varphi_k\sim \ln(2)/\alpha_k$. If the fields encounter this drop, inflation ends quickly and they roll to $-\infty$ in finite time; thus, the potential is unsuitable to describe the universe after inflation, but may be taken as a toy model during inflation, satisfying our demand for a separable Hubble parameter.

Given $H$ from (\ref{Hexample1}) we can compute $\delta_k$, $\gamma_k$ and $\beta_k$ from (\ref{deltak})-(\ref{betak}) to
\begin{eqnarray}
\delta_k&=&\left(\frac{H_0}{H(\tau)}\right)^2\frac{\alpha_k^2}{\mathcal{N}^2}e^{-2\alpha_k\varphi_k(\tau)}\,,\\
\gamma_k&=&-\alpha_k\sqrt{\delta_k}\,,\\
\beta_k&=&\alpha_k^2\delta_k=\gamma_k^2\,.
\end{eqnarray}
Since inflation requires $\delta=\sum_k\delta_k\ll 1/2$, we necessarily have $\delta_k \ll 1/2$ during inflation. However, for suitable large $\alpha_k$, that is for sharp cliffs in $V$, we can still achieve large $\gamma_k$ and $\beta_k$. As a consequence, large (but negative) NG are at least in principle possible, as we shall see below. We choose $t_c$ at the end of inflation, that is we determine $t_c$ by solving
\begin{eqnarray}
2\delta(t_c)=1
\end{eqnarray}
for $t_c$.
The volume expansion rate in (\ref{N}) from $t_c$ to $t_*$ can be integrated to
\begin{eqnarray}
N&=&\sum_k\left(\frac{\varphi_k^c-\varphi_k^*}{2\alpha_k}-\frac{e^{\alpha_k\varphi_k^c}-e^{\alpha_k\varphi_k^*}}{2\alpha_k^2}\right)\\
&=&\sum_k\left(\frac{1}{2\alpha_k^2}\ln\left(\frac{H^c\sqrt{\delta_k^c}}{H^*\sqrt{\delta_k^*}}\right)+\frac{H_0}{2\mathcal{N}H^*}\left(\frac{H^*}{H^c\gamma_k^c}-\frac{1}{\gamma_k^*}\right)\right)\,. \label{Nexample1}
\end{eqnarray}
In order to conform with current observations we need $N\sim 60$ and $\delta^c\sim 10^{-2}$, to give a scalar spectral index close to $1$.

So far, we have not specified the initial conditions, except for imposing one constraint, $N\sim 60$. In the two field case of \cite{Byrnes:2009qy} a set of initial conditions was found that yields a large (negative) $f_{NL}$, which can be generalized to
\begin{eqnarray}
\varphi_k^*=\frac{1}{\alpha_k}\ln\left(\left(N-\sqrt{\frac{m_k}{2\alpha_k^2}}\right)\frac{2\alpha_k^2}{\mathcal{N}}\right)\,, \label{init}
\end{eqnarray} 
where $m_k$ determines $\delta_k^c=1/(2m_k)$. If $\alpha_k>R\gg 1$ while $m_k<R^2$ with some large $R$,
the logarithm in (\ref{Nexample1}) can be neglected and, using $H_0\approx H^*$, one can write
\begin{eqnarray}
N&\approx&\sum_k\frac{1}{2\mathcal{N}}\left(\frac{H_0}{H^c\gamma_k^c}-\frac{1}{\gamma_k^*}\right)\,.
\end{eqnarray}
It is further possible \cite{Byrnes:2009qy} to provide a simple analytic estimate in the two field case, leading to the expectation that $f_{NL}=\mathcal{O}(\mathcal{N} \alpha_k/\sqrt{m_k})$ and $\tau_{NL}=\mathcal{O}(f_{NL}^2)$, see \cite{Byrnes:2009qy} for details (the full expressions for $\tau_{NL}$ where not computed there). We use the full analytic expression in (\ref{ns2}) for $n_s$, (\ref{fnle}) for $f_{Nl}^{(4)}$ and (\ref{tau2e})-(\ref{tau4e}) for $\Delta \tau_{NL}^{(i)}$. 

\subsection{Case-study $\mathcal{N}=2$\label{caseN=2}}
\begin{figure*}[tb]
\includegraphics[width=\textwidth,angle=0]{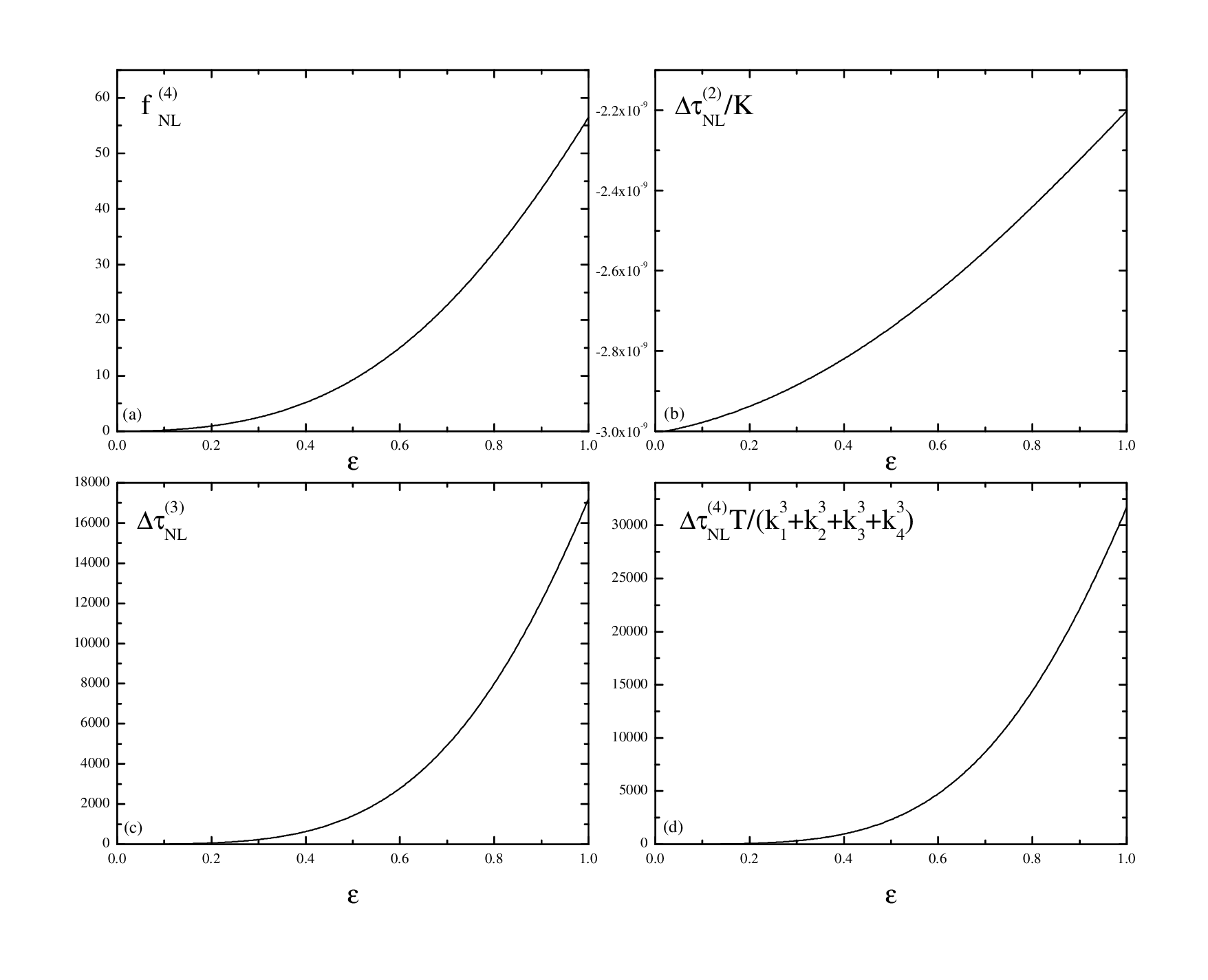}
   \caption{\label{pic:NG_Byrnes}
    We plot the non-linearity parameters $f_{NL}^{(4)}$, $\Delta \tau^{(2)}_{NL}$, $\Delta \tau^{(3)}_{NL}$ and $\Delta \tau^{(4)}_{NL}$ 
    from (\ref{fnle}) and (\ref{tau2e})-(\ref{tau4e}) over $\epsilon(t)=-\dot{H}/H^2$ for $\mathcal{N}=2$ and the initial conditions in (\ref{init}) with $\alpha_1=100$, $\alpha_2=20$, $m_1=6$ and $m_2=6/5$, as in \cite{Byrnes:2009qy}. The corresponding trajectory in field space is the forth from below in Fig.~\ref{pic:NG_Byrnes_inichanges}(a). $\Delta \tau^{(2)}_{NL}$ 
is unobservably small, while $\Delta \tau^{(3)}_{NL}$ and $\Delta \tau^{(4)}_{NL}$ roughly scale as $\left(f_{NL}^{(4)}\right)^2$. We have chosen $\epsilon$ as a time variable, since NG are only produced towards the end of inflation in this model  ($N(t_c)\approx 59.999$ when $\epsilon(t_c)=1$ and $N(\tilde{t})\approx 59.922$ when $\epsilon(\tilde{t})=0.1$).
}
\end{figure*}
Choosing the initial conditions according to (\ref{init}) with $\alpha_1=100$, $\alpha_2=20$, $m_1=6$ and $m_2=6/5$, we recover the result of \cite{Byrnes:2009qy}, that is $n_s-1\approx -0.033$ and $f_{NL}$ from Fig.~\ref{pic:NG_Byrnes} (a); we can also compute $\Delta \tau^{(i)}_{NL}$, see Fig.~\ref{pic:NG_Byrnes} (b)-(d). As expected, not only does $f_{NL}$ become large (but negative) once $2\delta=\epsilon$ approaches one, but also $\Delta\tau_{NL}^{(3)}$ and $\Delta\tau_{NL}^{(4)}$, both of which roughly scale as $\left(f_{NL}^{(4)}\right)^2$, as emphasized in \cite{Byrnes:2009qy}, while  $\Delta\tau_{NL}^{(2)}$ remains small.
 However, we would like to stress that the actual expressions differ by factors of order one from the simple estimates in \cite{Byrnes:2009qy} ($\Delta\tau_{NL}^{(3)}$ and $\Delta\tau_{NL}^{(4)}$ are roughly a factor of $3$ bigger than the estimate in \cite{Byrnes:2009qy} for this example \footnote{The approximation in \cite{Byrnes:2009qy} is valid if one 
of the fields gives the dominant contribution to $\zeta$ at all required orders. In the example above,
the approximation in \cite{Byrnes:2009qy} gets 
more accurate if $\alpha_2$ is decreased, approaching the $1\%$ 
level for $\alpha_2\sim 1$. We thank C.~Byrnes for discussions.}).
Since NG are only produced shortly before the end of inflation, we have chosen $\epsilon=-\dot{H}/H^2$ as a time variable ($N(t_{c})\approx 59.999$ when $\epsilon(t_{c})=1$ and $N(\tilde{t})\approx 59.922$ when $\epsilon(\tilde{t})=0.1$). 

\begin{figure*}[tb]
\includegraphics[width=\textwidth,angle=0]{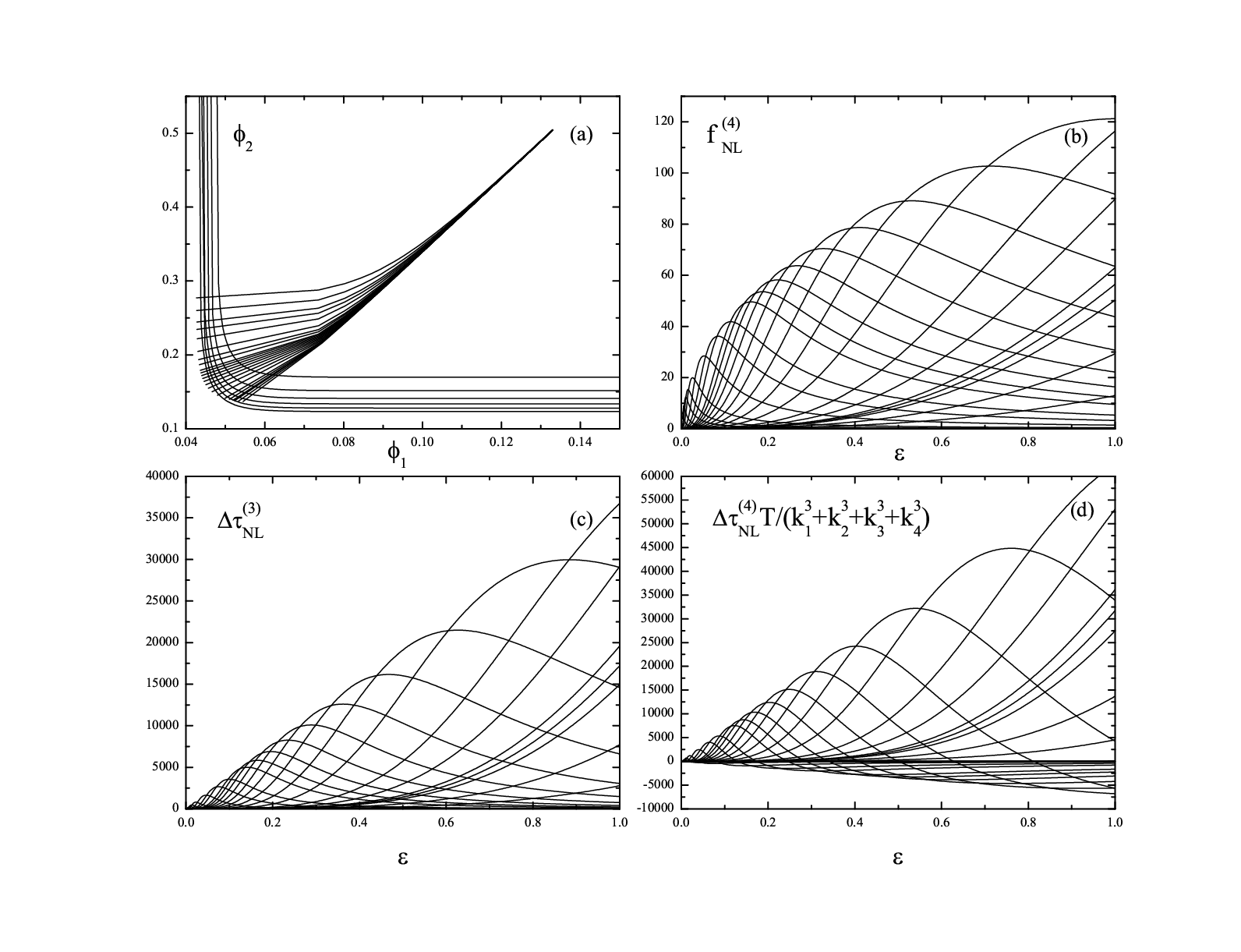}
   \caption{\label{pic:NG_Byrnes_inichanges}
We plot $f_{NL}^{(4)}$, $\Delta \tau^{(3)}_{NL}$ and $\Delta \tau^{(4)}_{NL}$ as in Fig.~\ref{pic:NG_Byrnes}, only changing the initial value for $\varphi_2$ slightly: instead of $\varphi_2^*$ from (\ref{init}), we use $c\varphi_2^*$ with $c=0.99999,0.999995,0.999999,1,1.000001,1.000005$ and $1.00001,\dots,1.0001,1.000125,1.00015,1.0002,\dots,1.0007,1.001$. In panel (a), we plot the corresponding trajectories, with increasing $c$ as the trajectories bend more. Superimposed are $V={\mbox{const}}$ lines for $V/H_0=n$ with $n=0,1,\dots,5$. Note the extreme sensitivity of NG to initial conditions, as well as their transient nature. Due to the latter, the non-linearty parameters should be evolved through the end-stages of inflation until fluctuations are imprinted onto radiation, to make predictions for the CMBR. In all cases $n_s-1\approx -0.033$  (effectively unchanged) and $\Delta \tau^{(2)}_{NL}$ remains unobservably small.}
\end{figure*}

It is instructive to alter the initial conditions slightly, see Fig.~\ref{pic:NG_Byrnes_inichanges}, where we vary $\varphi_2^*$: as soon as we deviate from (\ref{init}) in about $1$ part in $1000$, NG at the end of inflation become suppressed. The trajectories leading to large NG are the ones close to the line differentiating between trajectories bending towards the $\varphi_2$-axis or the $\varphi_1$-axis respectively, see Fig.~\ref{pic:NG_Byrnes_inichanges} (a).  Close to this repeller, isocurvature perturbations are unstable and grow; once the trajectory bends, they influence the adiabatic modes which become non-Gaussian to some extent. This is similar to the origin of non-Gaussianities in the new-ekpyrotic scenario, where two fields are supposed to roll along a crest in the potential; if the ekpyrotic phase is terminated by rolling off the crest, as in \cite{Koyama:2007if}, isocurvature modes are converted to adiabatic ones and as a byproduct NG are produced.

Interestingly, it can be seen in Fig.~\ref{pic:NG_Byrnes_inichanges} (b)-(d) that an increase in the non-linearity parameters appears to be a transient effect, which may or may not be imprinted onto radiation, depending on the nature of reheating which can take anywhere from a fraction of an e-fold in efficient preheating models \footnote{Note that preheating via parametric resonance is generically suppressed in multi-field models of inflation \cite{Battefeld:2008bu,Battefeld:2008rd,Battefeld:2009xw}, but tachyonic instabilities can still lead to a fast decay of the inflatons \cite{Battefeld:2009xw}.} to many e-folds in the case of the old theory or reheating (see \cite{Bassett:2005xm} for a review of (p)reheating). Since the enhancement and decay in Fig.~\ref{pic:NG_Byrnes_inichanges} takes place in a fraction of the last e-fold, it becomes crucial to follow the evolution of the non-linearity parameters through the decay of the inflatons to make unambiguous predictions. Note that the sign of $\Delta \tau^{(4)}_{NL}$ can change depending on the choice of $t_c$, see i.e.~Fig.~\ref{pic:NG_Byrnes_inichanges} (d), while $\Delta \tau^{(3)}_{NL}$ is always positive, in agreement with expectations from slow roll models, see i.e.~\cite{Huang:2009xa}. Unfortunately, the potential we focused on is unsuitable for $\epsilon> 1$. To apply our formalism to a toy model of (p)reheating, one needs to find an $H_k(\varphi_k)$ such that $V(\varphi_1,\dots,\varphi_{\mathcal{N}})$  possesses a global minimum after inflation. We postpone this interesting study to a future publication.   

We conclude that in this toy model NG are only produced for fine tuned initial conditions.

\subsection{Case-study $\mathcal{N}=2,\dots,6$\label{casengeneral}}
\begin{figure*}[tb]
\includegraphics[width=\textwidth,angle=0]{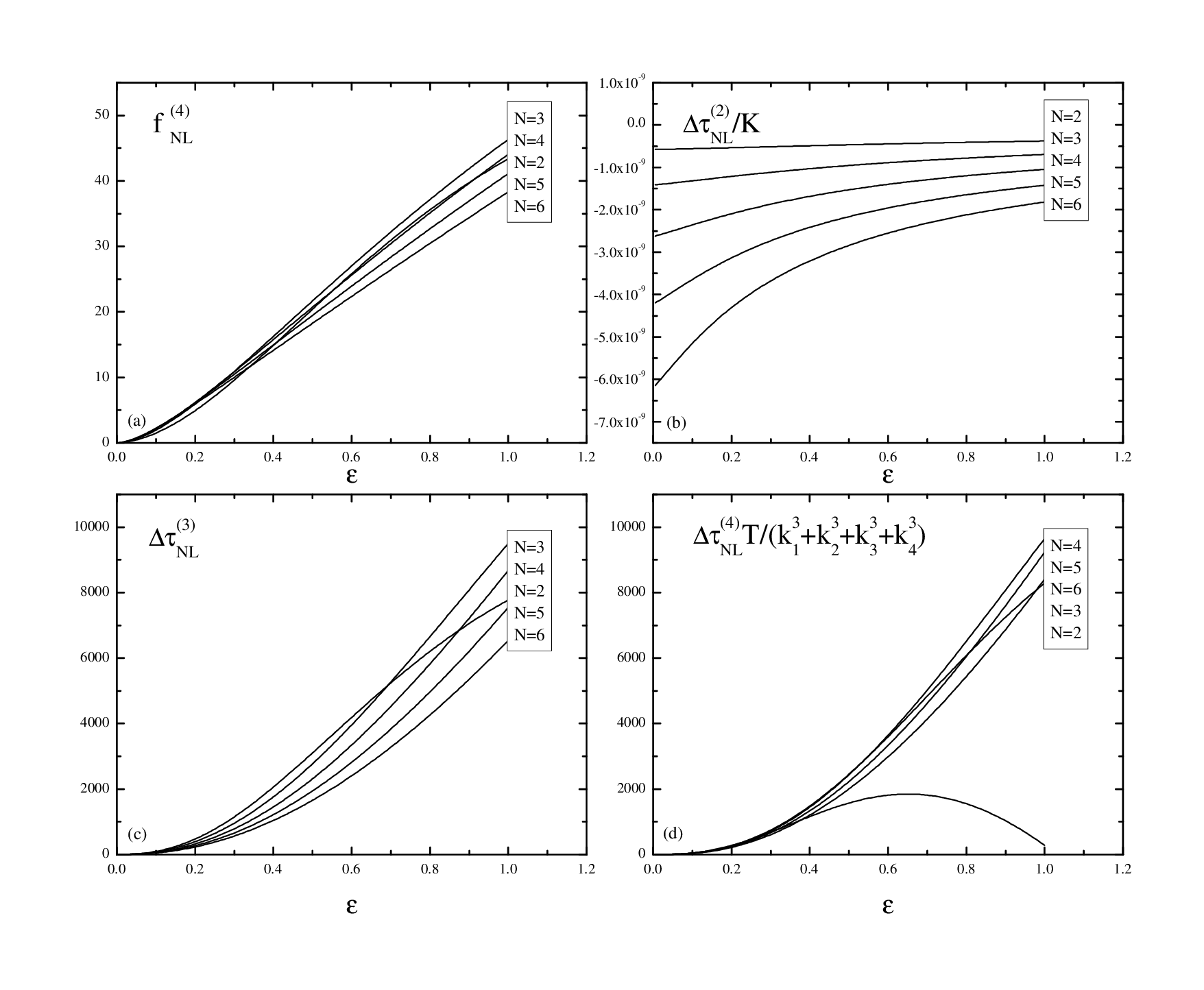}
   \caption{\label{pic:NG_Multi}
We plot $f_{NL}^{(4)}$ and $\Delta \tau^{(i)}_{NL}$ for $1=2,3,4$ from (\ref{fnle}) and (\ref{tau2e})-(\ref{tau4e}) over $\epsilon(t)=-\dot{H}/H^2$ for $\mathcal{N}=2\dots 6$ fields, with initial conditions according to (\ref{init}) with $\alpha_k=k\times 100/\mathcal{N}$ and $m_k=\mathcal{N}$, so  that observably large NG result.}
\end{figure*}
The $\mathcal{N}=2$-case is special in that symmetries in $A_{kl}$ and $A_{klm}$ cause cancellations in the non-linearity parameters that lead to the simple analytic expressions in  \cite{Byrnes:2009qy}, such as the one for $f_{NL}^{(4)}$ in (38) of \cite{Byrnes:2009qy}. Here, we stay close to the previous section's example, that is, we once again use initial conditions according to (\ref{fnle}), but we increase the number of fields from $\mathcal{N}=2\dots 6$. We further set $\alpha_k=k \times 100/\mathcal{N}$  and $m_k=\mathcal{N}$. As in Sec.~\ref{caseN=2}, the initial values are chosen to ensure observably large NG, albeit at the cost of fine tuning. In Fig.~\ref{pic:NG_Multi}, we plot $f_{NL}^{(4)}$ and $\Delta \tau^{(i)}_{NL}$ for $i=2,3,4$ using the full expressions in (\ref{fnle}) and (\ref{tau2e})-(\ref{tau4e}). The scalar spectral index ($n_s-1\approx -0.033$) is unaffected by increasing the number of fields.

The level of NG varies mildly if we increase the number of fields, but could be tuned to coincide at $\epsilon=1$ by small adjustments of the initial conditions. The largest difference in Fig.~\ref{pic:NG_Multi} is caused by raising $\mathcal{N}=2$ to $3$, simply because the change in the exponent of $H_k$ is largest ($\mathcal N=2$ has $\alpha_1=50$ and $\alpha_2=100$, while $\mathcal N=3$ has $\alpha_1\approx 33.3$, $\alpha_2\approx 66.6$ and $\alpha_3=100$), which directly impacts NG.

Fig.~\ref{pic:NG_Byrnes_inichanges} and \ref{pic:NG_Multi} illustrate the ambiguity problem of multi-field inflationary models. Whereas in the single-field case, the initial value, and thus all observables, are determined by the required e-folding number $N\sim 60$, there is already ambiguity in the two field case if only the two-point correlation function is considered; this may be lifted if NG are observed, since NG are usually undetectable in single field models, but can be larger if more fields are involved \footnote{Not all multi-field potentials offer the possibility to generate large NG; see i.e.~the first example in \cite{Byrnes:2009qy} for a case with unobservably small NG, (regardless of initial conditions), $\mathcal{N}$-flation \cite{Battefeld:2007en}, the examples in \cite{Battefeld:2006sz}, or the examples in this paper if the initial conditions are not fine tunned, among others.}, see Fig.~\ref{pic:NG_Byrnes_inichanges}. However, discriminating between $\mathcal{N}=2$ and $\mathcal{N}>2$ appears to be hopeless in the absence of a compelling reason for choosing one set of initial conditions over the other even if NG are observed. If one can choose starting points for the fields freely, any level of NG of a two-field case can be mimicked by a model with more fields. For instance, modest changes of the initial values in Fig.~\ref{pic:NG_Multi} similar to the ones in Fig.~\ref{pic:NG_Byrnes_inichanges} would allow for a tuning of the non-linearity parameters at $\epsilon=1$.

\section{Discussion and Conclusions \label{sec:conclusion}}
Based on the assumption of a separable Hubble parameter $H=\sum_kH_{k}(\varphi_k)$, we computed analytically Non-Gaussianities (NG), that is the non-linearity parameters characterizing the bi- and tri-spectrum, in multi-field inflationary models with an arbitrary number of fields and without using the slow roll approximation, the horizon crossing approximation, or a separable potential, extending the work of Byrnes and Tasinato \cite{Byrnes:2009qy}. Based on the $\delta N$-formalism, we derived analytic expression for $n_s$, $f_{NL}^{(4)}$ and $\Delta \tau^{(i)}_{NL}$ for $i=2,3,4$, which are easily computed once the background evolution of the fields is known.

To show the applicability of the formalism, we considered a simple exponential dependence of the Hubble parameter on the fields. This model has the advantage of being analytically solvable at the background level, but it comes at the price of being unrealistic once inflation comes to an end. If initial conditions are fine tuned, observably large NG can be produced towards the end of inflation, but the general prediction of this class of models are negligible NG. Further, even if NG are produced when $\epsilon=-\dot{H}/H^2\sim 1$, it remains to be seen if they are imprinted onto radiation after inflation, because the sudden increase of the non-linearity parameters appears to be a transient phenomenon.

We note that the formalism should be well suited for a computation of NG during (p)reheating, because the number of fields is not restricted, the slow roll approximation is not required and cross couplings between the fields are allowed. Indeed, in light of the example studied here, it appears mandatory to follow the non-linearity parameters through the era of inflaton decay, until fluctuations are imprinted onto radiation. We postpone a study of NG from (p)reheating to a forthcoming publication. 

\vspace{0.5cm}
\emph{Correction over Published Version:}\\  An ambiguity in the sign of $\sqrt{\delta_k}$ lead to a sign mistake for $f_{NL}$ in early versions of \cite{Byrnes:2009qy}. In this paper, all expressions are and have been correct to our knowledge, but Fig.~\ref{pic:NG_Byrnes}-\ref{pic:NG_Multi} contain wrong labels in the JCAP version and v1 and v2 on the arxiv (it should have been $f_{NL}^{(4)}$ instead of $-f_{NL}^{(4)}6/5$). The labels are corrected in this version v3. Further, whenever we remark that $f_{NL}$ can be large, we added that its sign is negative and thus already strongly constrained by observations. We thank C.~Byrnes and G.~Tasinato for alerting us to this mistake.

\begin{acknowledgments}
We would like to thank C.~Byrnes and G.~Tasinato for comments. T.B. is supported by the Council on Science and Technology at Princeton University and is thankful for hospitaliy at the HIP and the APC. D.~B was supported by the EU EP6 Marie Curie Research and Training Network 'UniverseNet' (MRTN-CT-2006-035863) and is thankful for hospitality at Princeton University, the APC, and thanks Kari Enqvist for support. This paper is dedicated to the memory of Erlinda Balmaceda (1927-2009).

\end{acknowledgments}

\end{document}